\documentclass[sigconf]{acmart}
\usepackage{adjustbox}
\usepackage{makecell}
\usepackage{bm,multirow}
\usepackage{arydshln}

\AtBeginDocument{%
  }

\setcopyright{acmcopyright}
\copyrightyear{2023}
\acmYear{2023}
\acmDOI{XXXXXXX.XXXXXXX}

\acmConference[Preprint]{Make sure to enter the correct conference title from your rights confirmation emai}{Under View}{}
\begin{document}

\title{ITportrait: Image-Text Coupled 3D Portrait Domain Adaptation}

\vspace{-3mm}
\author{Xiangwen Deng, Yufeng Wang, Yuanhao Cai, Jingxiang Sun, Yebin Liu, and Haoqian Wang}
\affiliation{%
  \institution{Tsinghua University}
  \country{China}
}

\renewcommand{\shortauthors}{Deng et al.}
%
\begin{abstract}
Domain adaptation of 3D portraits has gained more and more attention.
However, the transfer mechanism of existing methods is mainly based on vision or language, which ignores the potential of vision-language combined guidance. 
In this paper, we propose an Image-Text multi-modal framework, namely Image and Text portrait (ITportrait), for 3D portrait domain adaptation. ITportrait relies on a two-stage alternating training strategy. In the first stage, we employ a 3D Artistic Paired Transfer (APT) method for image-guided style transfer. APT constructs paired photo-realistic portraits to obtain accurate artistic poses, which helps ITportrait to achieve high-quality 3D style transfer. In the second stage, we propose a 3D Image-Text Embedding (ITE) approach in the CLIP space. ITE uses a threshold function to self-adaptively control the optimization direction of images or texts in the CLIP space. Comprehensive experiments prove that our ITportrait achieves state-of-the-art (SOTA) results and benefits downstream tasks. All source codes and pre-trained models will be released to the public. 

\end{abstract}


\begin{teaserfigure}
\begin{center}
  \includegraphics[width=0.95\textwidth]{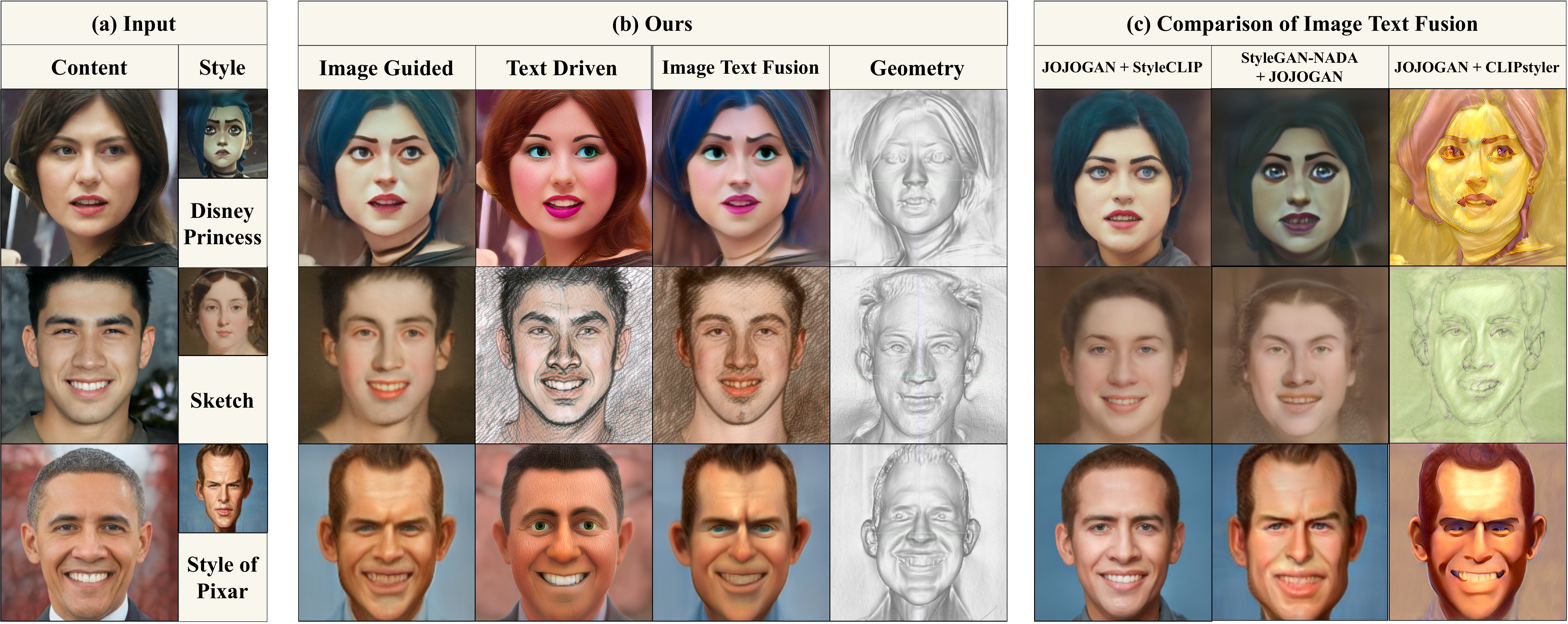}
  \end{center}
  \vspace{-2mm}
  \caption{(a) Input style image and text description. (b) Our ITportrait can support Image-guided style transfer, Text-driven editing, and Image-Text coupled domain adaptation.
  (c) Directly applying style transfer (StyleGAN-NADA~\cite{stylegannada}, JOJOGAN~\cite{JoJoGAN0}) and text editing (StyleCLIP~\cite{Styleclip2}, CLIPstyler~\cite{CLIPstyler0}) methods achieves inferior Image-Text coupled domain adaptation results.} 
  \label{fig:Teaser}
  \vspace{2mm}
\end{teaserfigure}


\maketitle

\section{Introduction}

Artistic portraits have many applications~\cite{Resolution2,Pastiche,Your3dEmoji22,DDSG16,DDSG5} in our daily lives, especially in industries related to animation, art, and the metaverse. As shown in Fig. \ref{fig:Teaser}, artistic portraits can be regarded as a portrait domain adaptation task, which refers to transforming the artistic style, cross-species identity, and expression shape change. The current domain adaption methods are mainly divided into two categories: guided by the vision-based method (artistic-image \cite{Pastiche,jojogan21,JoJoGAN0}), or guided by the language-based method (text-description \cite{Styleclip2,cfclip0}). Combining image-guided and text-driven guidance can not only transfer the precise and detailed style of the reference image but also have the text-driven flexible editing ability. Therefore, Image-Text coupled guidance has better style control-ability and artistic merit \cite{StyleCLIPDraw0,styleclipdrawFrans}. However, the potential of Vision-Language (Image-Text) multi-modal guidance is under-explored.

\vspace{4mm}

Mixing styles from the artistic images and the text description is challenging. To achieve image-text coupled domain adaptation, we need to first implement the transfer of artistic image style and text description, respectively. Nevertheless, as shown in Fig.~\ref{fig:Teaser}, when directly using the method of style transfer and text editing successively, the style of the previous stage will be lost. In addition, compared with 2D GAN-based methods, 3D GANs~\cite{3DAvatarGAN,11StyleNerf,40Ide3d,55Cips3d,piGAN} have more advantages in domain adaptation tasks due to the ability of multi-view consistency synthesis. However, achieving portrait domain adaptation on 3D GAN also exacerbates the difficulty. The reasons are twofold: 1) From the image-guided perspective, it is challenging for the existing methods~\cite{3DAvatarGAN,dr3d,Your3dEmoji0} to achieve 3D high-quality style transfer. Because the pose of the artistic style reference image is tough to estimate. 2) From the text-driven aspect, previous 3D portrait text-driven methods~\cite{sun2022next3d,40Ide3d} may cause geometric collapse (Fig~\ref{fig:ablation2}) when supervised in a single view \cite{stylegannada,CLIPstyler0}.

 To cope with the above problems, we propose a multi-modal framework, ITportrait (short for "Image and Text"), which supports domain adaptation for 3D portraits jointly guided by images and texts. Inspired by CLIP's ability to encode images and texts, we consider embedding the image and text feature in the CLIP space to mix image and text styles simultaneously. However, using a single image embedding in the CLIP space as guidance will lead to overfitting~\cite{stylegannada}. Hence, we additionally train a $\mathcal{G}_{3d}^{s}$ to generate sufficient stylized samples, which will be embedded in the CLIP space as image guidance to prevent overfitting. Subsequently, we propose a two-stage alternating training strategy. In the first stage, we design a 3d portrait style transfer method (APT). We construct paired photo-realistic portrait images, which can help to obtain accurate pose estimation of art-reference images. By this means, a high-quality one-shot style transfer can be realized.
In the second stage, we propose an Image-Text Embedding (ITE) strategy that includes a text-guided direction and an image-guided direction in the CLIP space. More specifically, the image guidance transfers the global style while the text guidance edits the local portraits. We employ a threshold function to control the direction of domain adaption in the CLIP space. Furthermore, a 3D multi-view augment strategy is proposed to prevent geometry collapse and improve the rendering quality. 
 
 To sum up, our contributions are listed as follows:

\begin{itemize}
\item We propose a multi-modal domain adaption framework, ITportrait. To the best of our knowledge, this work is the first attempt to explore the potential of the Image-Text coupled domain adaption for 3D portraits. 
\item We customize an Image-Text embedding approach, namely ITE, to self-adaptively control the fusion of image and text guidance in the CLIP space. ITE not only employs an artistic paired transfer method APT to avoid one-shot stylization overfitting but also constructs 3D multi-view supervision to improve the geometric quality of the text-driven editing.
\item Our ITportrait boosts the downstream application of 3D-aware one-shot portrait tasks including image-guided stylization, text-driven manipulation, and Image-Text coupled domain adaption. ITportrait also pushes the frontier of view-consist editing for photo-realistic and art-drawing portraits.
\end{itemize}
\vspace{0.6mm}

\section{Related Work}

\subsection{Artistic Portrait Generation}
 The artistic portrait is generally generated by GANs. Many works in the 2D vision field are based on StyleGAN~\cite{stylegan2}. For instance, Toonify~\cite{Resolution2} achieves style transfer by exchanging the layers of StyleGAN. BlendGAN~\cite{jojogan21} performs style transfer by training an MLP and injecting style into StyleGAN. JOJOGAN~\cite{JoJoGAN0} constructs a paired dataset to avoid overfitting. Recently, DualstyleGAN~\cite{Pastiche} yields high-quality style transfer by training a dual-path StyleGAN. With the development of 3DGAN~\cite{1EG3D,40Ide3d,11StyleNerf}, more and more works are studying 3D artistic portrait generation. For example, Dr3d~\cite{dr3d} and 3DAvatarGAN~\cite{3DAvatarGAN} are trained from drawing datasets~\cite{Resolution2, Pastiche} to achieve domain adaption. However, these large-scale corresponding domain datasets are tedious and labor-intensive to obtain. The closest work to us is Your3dEmoji~\cite{ Your3dEmoji0}, which can complete a one-shot 3D style transfer. Nevertheless, Your3dEmoji cannot obtain the pose of artistic reference images, which leads to time-consuming transfer learning and causes a poor effect. Besides, IDE-3D~\cite{40Ide3d} and Next3d~\cite{sun2022next3d} use the CLIP-based method (StyleGAN-NADA~\cite{stylegannada}) to generate artistic portraits. However, these CLIP-based augment methods (\emph{i.e.,} perspective augments~\cite{styleclipdrawFrans,CLIPstyler0}) are all designed for 2D vision, which only considers the front-view supervision for 3D portraits. It may lead to poor generation performance for the side of the portrait or geometry collapse, as shown in Fig. \ref{fig:ablation2}. In contrast, our APT can predict an accurate artistic pose to improve the quality of image-based one-shot style transfer. And we construct 3D multi-view samples in CLIP space to enhance the portrait render quality.

\begin{figure*}[tp]
\begin{center}
\includegraphics[width=0.95 \linewidth]{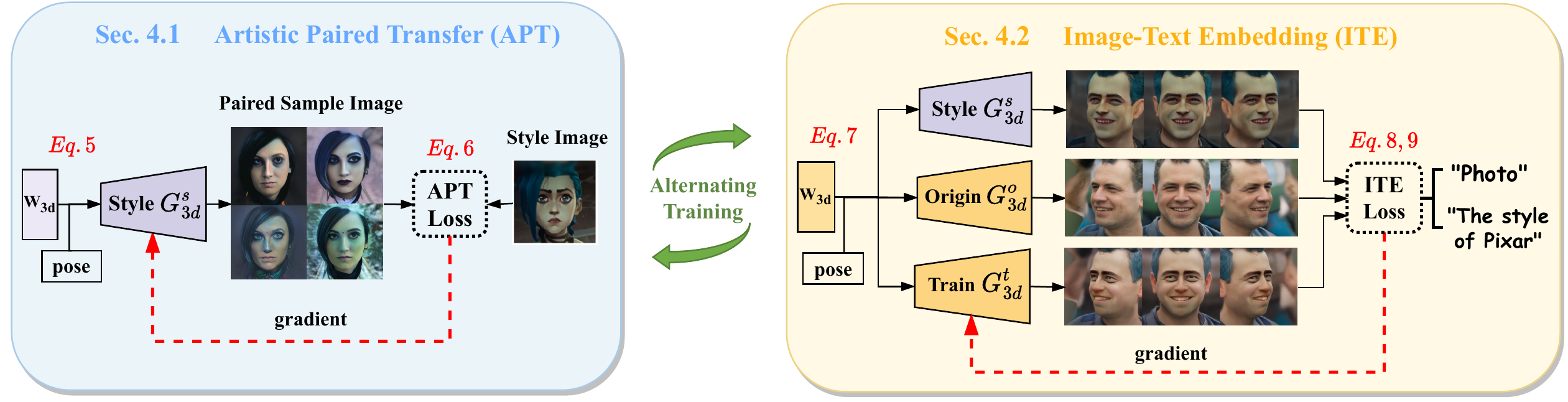}
\end{center}
   \caption{Overview of our ITportrait pipeline. Our method adopts a two-stage alternating training strategy. The first stage in \textcolor{blue}{blue} is image-guided one-shot stylization, as described in Sec \ref{section_image}. The gradient of Eq.~\eqref{equation_jojo} is used for training $\mathcal{G}_{3d}^{s}$. The second stage in \textcolor{yellow}{yellow} is Image-Text fusion in the CLIP space, as detailed in Sec \ref{section_mix}, where $\mathcal{G}_{3d}^{s}$ comes from the image-guided stylization stage, $\mathcal{G}_{3d}^{o}$ and $\mathcal{G}_{3d}^{t}$ comes from the pre-trained real photographs domain (\emph{i.e.,} FFHQ~\cite{stylegan2}). The gradient of Eq.~\eqref{clip_text} is used to train $\mathcal{G}_{3d}^{t}$. Alternately training the above two stages can achieve a high-quality Image-Text fusion adaptation effect.}
\label{fig_pipleline_4}
\end{figure*}

\subsection{Image-Text Based Domain Adaption}
    Portrait domain adaptation can be divided into twofold: image-guided methods and text-driven methods. The earliest image-guided method can be regarded as image-to-image translation task~\cite{nada4,Styleclip2,dualstylegan32,dualstylegan38,dualstylegan39}. The current image-guided methods like DynaGAN~\cite{kim2022dynagan}, and StyleDomain~\cite{alanov2022styledomain} transfer styles by modifying the structure of StyleGAN~\cite{Training2,nada21,stylegan2}. The main disadvantage of imaged-guided transfer is that it only supports to transfer of the global style, but the local editing of portraits is less flexible than the text-based method. For text-guided transfer, the CLIP model~\cite{radford2021learning} has become the mainstream method of text-driven synthesis since it can learn a joint embedding space of images and text. For example, StyleCLIP~\cite{Styleclip2} achieves high-fidelity manipulation by exploring the latent space of StyleGAN. CF-CLIP~\cite{cfclip0} can support better text-driven face controllable editing. The main disadvantage of text-driven methods is that although the local editing of portraits is convenient, the global cross-domain transfer of portraits is less controllable than image-based methods. Recently, researchers proposed some Image-Text couple methods for domain adaption. StyleCLIPDraw~\cite{StyleCLIPDraw0} and CLIPDraw~\cite{styleclipdrawFrans} propose an Image-Text coupling generator method, proposed an Image-Text coupled generator approach. However, their approach is for generation from scratch rather than further manipulation of specified images. StyleGAN-NADA~\cite{stylegannada} proposes a direction-loss of image and text in the CLIP space. Although StyleGAN-NADA can integrate images and text style, it requires many art reference images as guidance. Otherwise, single image guidance in the CLIP space will cause the gender and expression of the content image to change, leading to overfitting and poor style transfer results. In contrast, We constructed an alternating training method. Additionally, training a $\mathcal{G}_{3d}^{s}$ to generate many style reference samples solved the overfitting problem of the one-shot image embedding with CLIP space.


\section{Preliminaries}
\noindent{\bf EG3D.} We first briefly review the network architecture of a SOTA 3D network, EG3D~\cite{1EG3D}. Our generator $\mathcal{G}_{3d}^{s}$, $\mathcal{G}_{3d}^{t}$ and $\mathcal{G}_{3d}^{o}$ are fine-tuned on EG3D. Specifically, EG3D starts with randomly sampled GAN latent codes, and then a feature generator based on StyleGAN~\cite{stylegan2} converts latent codes into 2D features and maps them into 3D tri-planes. An MLP decoder predicts features for 3D point projections on tri-planes to generate color and density. Finally, volume rendering is used to generate an image on the orientation of the camera pose. This process can be formulated as follows:

\begin{equation} 
\begin{aligned}
 \mathbf{I}=\mathcal{G}_{3d}(\mathbf{W}_{3d},\mathbf{P};\theta),  
\label{equation_eg3d}
\end{aligned}
\end{equation}

\noindent where $\mathbf{W}_{3d}\in \mathbb{R}^{1 \times 14 \times 512}$ is the lantent code. $\mathbf{P}\in \mathbb{R}^{ 1 \times 25}$ is the camera pose. $\theta$ is the weight parameters of the EG3D generator $\mathcal{G}_{3d}$. $\mathbf{I} \in \mathbb{R}^{H\times W \times N}$ are the generated images.

\noindent {\bf CLIP-guided Loss.} OpenAI proposes CLIP~\cite{radford2021learning}, a high-performance text-image embedding model trained on 400 million text-image pairs. It consists of two encoders that encode images and text into 512-dimensional embeddings, respectively. A later work, StyleGAN-NADA~\cite{stylegannada}, designs a direction CLIP loss to align the CLIP space directions between the source and target text-image pairs as 

\begin{equation}
 \begin{aligned}
\Delta{T}=\mathcal{E}_T(\boldsymbol{T}_{target})-\mathcal{E}_T(\boldsymbol{T}_{source}), \\
 \Delta{I}=\mathcal{E}_I(\mathbf{I}_{train})-\mathcal{E}_I(\mathbf{I}_{source}), \\
\boldsymbol{L}=1 - \frac{\Delta{T} \cdot \Delta{I}}{\left\lvert{\Delta{T}}\right\rvert  \left\lvert{\Delta{I}}\right\rvert }, 
  \label{prim_clip}
 \end{aligned}
\end{equation}

 \noindent where, $\mathcal{E}_I$ and $\mathcal{E}_T$ are the image and text encoder of CLIP, $ \boldsymbol{T}_{target}$ and $\boldsymbol{T}_{source}$ denote the text and input content of the style, respectively. When we use natural images as content, the $\boldsymbol{T}_{source}$ is set to "photo". $\mathbf{I}_{source}$ refers to the source image. $\mathbf{I}_{train}$ represents the manipulated images. $\boldsymbol{L}$ indicates the direction CLIP loss.


\section{Method} 
In this section, we describe the proposed ITportrait framework. Our goal is to achieve Image-Text coupled domain adaption of 3D portraits. We consider embedding the image and text features in the CLIP space. Yet, using a single image as guidance will lead to overfitting~\cite{stylegannada}. Hence, we design an alternating training approach as shown in Fig. \ref{fig_pipleline_4}. Specifically, the alternating training approach includes two stages: the one-shot style transfer stage in Sec. \ref{section_image} and the Image-Text fusion stage in Sec. \ref{section_mix}.
The one-shot style transfer (APT) stage generates sufficient stylization images to prevent portraits from overfitting. The Image-Text fusion stage (ITE) embeds stylization images and text descriptions in the CLIP space for Image-Text fusion adaptation. Eventually, ITportrait can gradually achieve Image-Text coupled domain adaption by alternately training the two stages.

\subsection{Image-guided One-shot Stylization}\label{section_image}

In this section, we propose an artistic paired transfer method (APT) to achieve image-guided 3D style transfer, as shown in Fig. \ref{fig_pipleline_4} with blue color. Specifically, we first propose a GAN inversion method to get the $\mathbf{W}_{3d}$ and pose $\mathbf{P}$ of artistic images. Then we use $\mathbf{W}_{3d}$ perturbations to construct paired datasets for stylization. 

\begin{figure}[tp]
\begin{center}
\includegraphics[width=0.95 \linewidth]{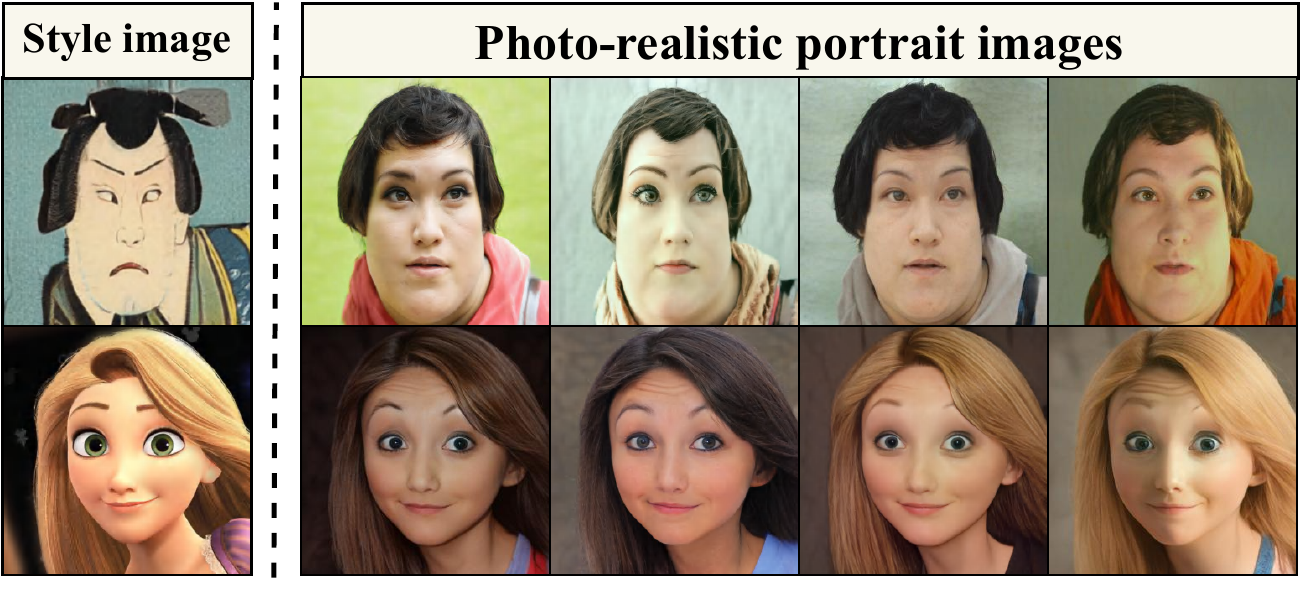}
\end{center}
   \caption{The photo-realistic portrait images ($\mathcal{G}_{2d}(\mathbf{W}'_{2d};\theta)$) are well-aligned with the artistic style images ($\mathbf{I}_{style}$). Thus, APT can obtain accurate pose estimation $\mathbf{P}$ from photo-realistic portrait images as an alternative to the artistic style image.}
\label{fig_ph_p}
\end{figure}

To begin with, we propose an artistic GAN inversion approach in Eq.~\eqref{equation_w2d} to predict the $\mathbf{W}_{3d}$ and pose $\mathbf{P}$. In practice, randomly initialized pose $\mathbf{P}$~\cite{ko20233d,yin20223d} can easily lead to optimization failures. Therefore, we need to obtain the artistic image's accurate initial pose $\mathbf{P}$. As shown in Fig. \ref{fig_ph_p}, we observe the existing 2D GAN inversion method~\cite{Designing2} can construct a well-aligned real photograph from the artistic reference image. Although estimating the pose of artistic images is difficult, the portrait in the photo-realistic domain is easy to estimate. Hence, our optimization-based GAN inversion approach to obtain the $\mathbf{W}_{3d}$ and $\mathbf{P}$ can be formulated as follows:

\begin{equation} 
\begin{aligned}
\mathbf{W}'_{2d}=\alpha \times \mathbf{W}_{2d} + (1-\alpha) \times \mathcal{M}(z), \\
\mathbf{P}_{init} = \mathcal{P}(\mathcal{G}_{2d}(\mathbf{W}'_{2d};\theta)), 
\label{equation_w2d}
\end{aligned}
\end{equation}

\begin{equation} 
\begin{aligned} 
\mathbf{W}_{3d}^{*}, \mathbf{P}^{*} =\mathop{\arg\min}\limits_{\mathbf{W}_{3d},\mathbf{P}}\mathcal{L}(\mathbf{I}_{style},\mathcal{G}_{3d}^{s}(\mathbf{W}_{3d},\mathbf{P}_{init} ;\theta)),
\label{equation_w3d}
\end{aligned}
\end{equation}

\noindent where $\mathbf{W}_{2d}\in \mathbb{R}^{1 \times 18 \times 512}$ is obtained from style image by the 2D encoder-based GAN inversion method Restyle~\cite{Designing2}. $z\in \mathbb{R}^{1 \times 512}$ is the randomly sampled noise. $\mathcal{M}(·)$ denotes the style mapping layers of StyleGAN~\cite{stylegan2}. $\alpha$ is a hyperparameter and set to $0.2$. $\mathbf{W}'_{2d}\in \mathbb{R}^{1 \times 18 \times 512}$ is obtained from perturbing 13-18 mapping layers of $\mathbf{W}_{2d}$. The pre-trained parameter $\theta$ of $\mathcal{G}_{2d}$ is fixed. $\mathcal{G}_{2d}(\mathbf{W}'_{2d};\theta)$ refers to the aligned photo-realistic images get from StyleGAN. Thus, we can employ an off-the-shelf estimator~\cite{deng2019accurate} $\mathcal{P}$ to predict the pose $\mathbf{P}_{init}\in \mathbb{R}^{1 \times 25}$ from the real photograph as an alternative to the artistic reference image.
Then, the pose $\mathbf{P}_{init}$ is used as the initialization in Eq.~\eqref{equation_w3d}. $\mathbf{I}_{style}\in \mathbb{R}^{H\times W \times N}$ is the style reference image. $\mathcal{G}_{3d}^{s}(\mathbf{W},\mathbf{P}_{init};\theta)$ refers to the align-paired images we get from EG3D. Loss $\mathcal{L}$ refers to $\mathcal{L}_{2}$, $\mathcal{L}_{LPIPS}$~\cite{jojogan39}, $\mathcal{L}_{ID}$, and $\mathcal{L}_{depth}$. This way, our optimize-base artistic GAN inversion approach in Eq.~\eqref{equation_w3d} can obtain the $\mathbf{W}_{3d}\in \mathbb{R}^{1 \times 14 \times 512}$ and $\mathbf{P}\in \mathbb{R}^{1 \times 25}$ of the artistic image. Please refer to our supplementary materials for more details about artistic GAN inversion (Eq.~\eqref{equation_w2d} and Eq.~\eqref{equation_w3d}).

After artistic GAN inversion, we start 3DGAN style transfer. Following the same way of 2D style transfer method JOJOGAN~\cite{JoJoGAN0}, our 3D style transfer process can be formulated as follows:

\begin{equation} 
\begin{aligned}
 \mathbf{W}'_{3d}=\beta \times \mathbf{W}_{3d} + (1-\beta) \times \mathcal{M} (z), \\
 \mathbf{I}_{gen}=\mathcal{G}_{3d}^{s}(\mathbf{W}'_{3d},\mathbf{P};\theta),  
\label{equation_jojo_w}
\end{aligned}
\end{equation}

\begin{equation} 
\begin{aligned}
 \theta^{*} =\mathop{\arg\min}\limits_{\theta}\mathcal{L}(\mathbf{I}_{style}, \mathbf{I}_{gen}),
\label{equation_jojo}
\end{aligned}
\end{equation}

\noindent where $\mathbf{W}_{3d}\in \mathbb{R}^{1 \times 14 \times 512}$ is given from Eq.~\eqref{equation_w3d}. $\mathcal{M}(·)$ denotes the style mapping layers of EG3D. $z\in \mathbb{R}^{1 \times 512}$ is the randomly sampled  noise. $\beta \in (0,0.2)$ controls the degree of stylization. The interpolated $\mathbf{W}'_{3d}\in \mathbb{R}^{1 \times 14 \times 512}$ is obtained from perturbing 9-13 mapping layers of $\mathbf{W}_{3d}$. The loss $\mathcal{L}$ refers to $LPIPS$. In each training epoch, we perturb 9-13 mapping layers of $\mathbf{W}_{3d}$ to get the interpolated $\mathbf{W}'_{3d}\in \mathbb{R}^{1 \times 14 \times 512}$. Subsequently, we use $\mathcal{G}_{3d}^{s}$ to utilize these $\mathbf{W}'_{3d}$ to generate aligned photo-realistic paired datasets $\mathbf{I}_{gen}\in \mathbb{R}^{H\times W \times N}$. As shown in Fig. \ref{fig_pipleline_4} blue part, these paired photo-realistic datasets $\mathbf{I}_{gen}$ will be supervised by style reference image $\mathbf{I}_{style}$ to make the 3DGAN style transfer.  We only fine-tune the parameters $\theta$ of $\mathcal{G}_{3d}^{s}$. With these techniques, we enable the EG3D generator $\mathcal{G}_{3d}^{s}$ to achieve high-quality one-shot stylization. 

\subsection{Image-Text Fusion in the CLIP Space} \label{section_mix}
In this section, our goal is to apply Image-Text embedding (ITE) manipulation to 3D portraits through the pre-trained text-image embedding model CLIP~\cite{radford2021learning}, as depicted in Fig. \ref{fig_pipleline_4} with yellow color. First, 3D multi-view images $\mathbf{I}_{style}$, $\mathbf{I}_{train}$ and $\mathbf{I}_{source}$ are generated by $\mathcal{G}_{3d}^{s}$, $\mathcal{G}_{3d}^{t}$ and $\mathcal{G}_{3d}^{o}$. Then, the 3D multi-view images are fed into the CLIP space for mixing with a text description. Finally, we propose a threshold function and CLIP-direction loss to control the fusion direction self-adaptively. Specifically, ITE constructs $\mathbf{I}_{style}\in \mathbb{R}^{H\times W \times N}$, $\mathbf{I}_{train}\in \mathbb{R}^{H\times W \times N}$, and $\mathbf{I}_{source}\in \mathbb{R}^{H\times W \times N}$ as

\begin{equation}
\begin{small}
 \begin{aligned}
\mathbf{I}_{style}=\mathcal{G}_{3d}^{s}(\mathbf{W}_{3d},\mathbf{P}_{m};\theta), \\
\mathbf{I}_{train}=\mathcal{G}_{3d}^{t}(\mathbf{W}_{3d},\mathbf{P}_{m};\theta), \\
 \mathbf{I}_{source}=\mathcal{G}_{3d}^{o}(\mathbf{W}_{3d},\mathbf{P}_{c};\theta), \\  
  \mathbf{P}_{m}= \sum_{i}^{n}\mathbf{P}_{3d}, \quad
\mathbf{P}_{c}=\sum_{i}^{n}\mathbf{P}_{2d},
 \label{MIX_clip}
 \end{aligned}
 \end{small}
\end{equation}

\noindent where $\mathbf{P}_{m}\in \mathbb{R}^{1 \times 25}$ refers to 3D multi-view samples. $\mathbf{P}_{c}\in \mathbb{R}^{1 \times 25}$ refers to 2D canonical view samples. The parameter $\theta$ of $\mathcal{G}_{3d}^{t}$ and $\mathcal{G}_{3d}^{o}$ in Eq.~\eqref{MIX_clip} is from the original EG3D. The parameter $\theta$ of $\mathcal{G}_{3d}^{s}(\mathbf{W}_{3d},\mathbf{P}_{m};\theta)$ is training in Sec. \ref{section_image}. $\mathbf{I}_{style}$ is an stylization image generated from $\mathcal{G}_{3d}^{s}(\mathbf{W}_{3d},\mathbf{P}_{m};\theta)$. Then, as emphasized in Eq.~\eqref{prim_clip} in the preliminary, ITE implements the 3D Image-Text embedding strategy by using CLIP-space direction as follows:

\begin{equation}
 \begin{aligned}
\Delta{T}=\mathcal{E}_T(\boldsymbol{T}_{target})-\mathcal{E}_T(\boldsymbol{T}_{source}), \\
\Delta{I}=\mathcal{E}_I(\mathbf{I}_{style})-\mathcal{E}_I(\mathbf{I}_{source}), \\
\Delta{IT}=\mathcal{E}_I(\mathbf{I}_{train})-\mathcal{E}_I(\mathbf{I}_{source}), \\
   \boldsymbol{L}_{I}=1 - \frac{\Delta{IT} \cdot \Delta{I}}{\left\lvert{\Delta{IT}}\right\rvert  \left\lvert{\Delta{I}}\right\rvert }, \\
    \boldsymbol{L}_{T}=1 - \frac{\Delta{IT} \cdot \Delta{T}}{\left\lvert{\Delta{IT}}\right\rvert  \left\lvert{\Delta{T}}\right\rvert }, \\
\boldsymbol{L}_{IT}=\gamma \cdot \boldsymbol{L}_{I}+ (1-\gamma) \cdot  \boldsymbol{L}_{T}, \\
 \end{aligned}
 \label{clip_text}
\end{equation}

\begin{figure}[tp]
\begin{center}
\includegraphics[width=0.95 \linewidth]{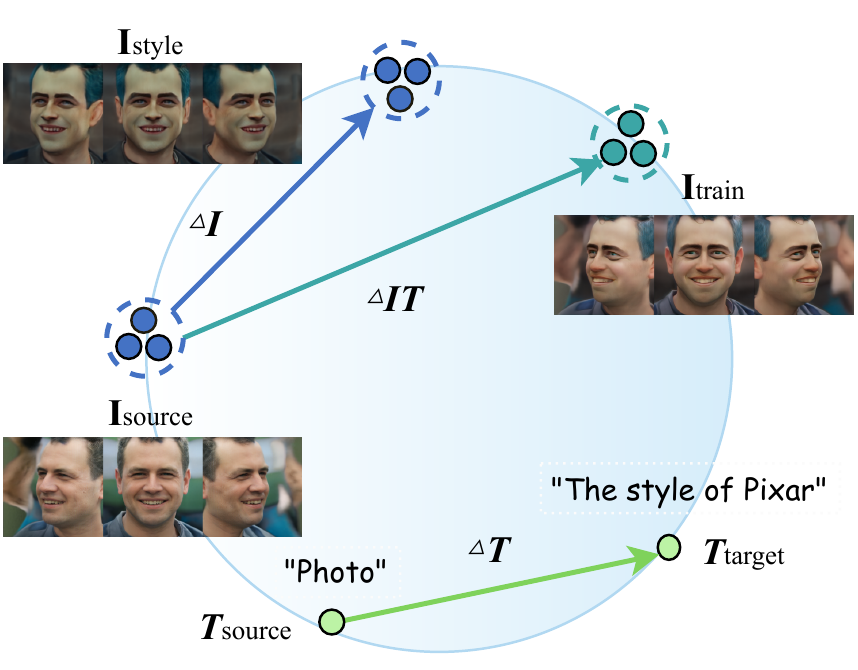}
\end{center}
\caption{Illustration of Image-Text coupled domain adaption in our CLIP-direction
loss in Eq.~\eqref{clip_text}. The generated images $\mathbf{I}_{style}$, $\mathbf{I}_{train}$ and $\mathbf{I}_{source}$ from generator $\mathcal{G}_{3d}^{s}$, $\mathcal{G}_{3d}^{t}$ and $\mathcal{G}_{3d}^{o}$ are embedded in the CLIP-space (Eq.~\eqref{MIX_clip}). The 3D-view images $\mathbf{I}_{source}$ and $\mathbf{I}_{train}$ construct the $\Delta{IT}$. The 3D-view image $\mathbf{I}_{source}$ and $\mathbf{I}_{style}$ construct the $\Delta{I}$. The text description constructs the $\Delta{T}$. For Image-Text coupled manipulation of the portrait
in Sec.~\ref{section_mix}, we require $\Delta{IT}$ to be parallel between $\Delta{T}$ and $\Delta{I}$. We self-adaptively control the parallelism according to the weight $\gamma$ in Eq.~\eqref{clip_para2}.}
\label{fig_clip2}
\end{figure}
\begin{figure}[tp]
\begin{center}
\includegraphics[width=0.9\linewidth]{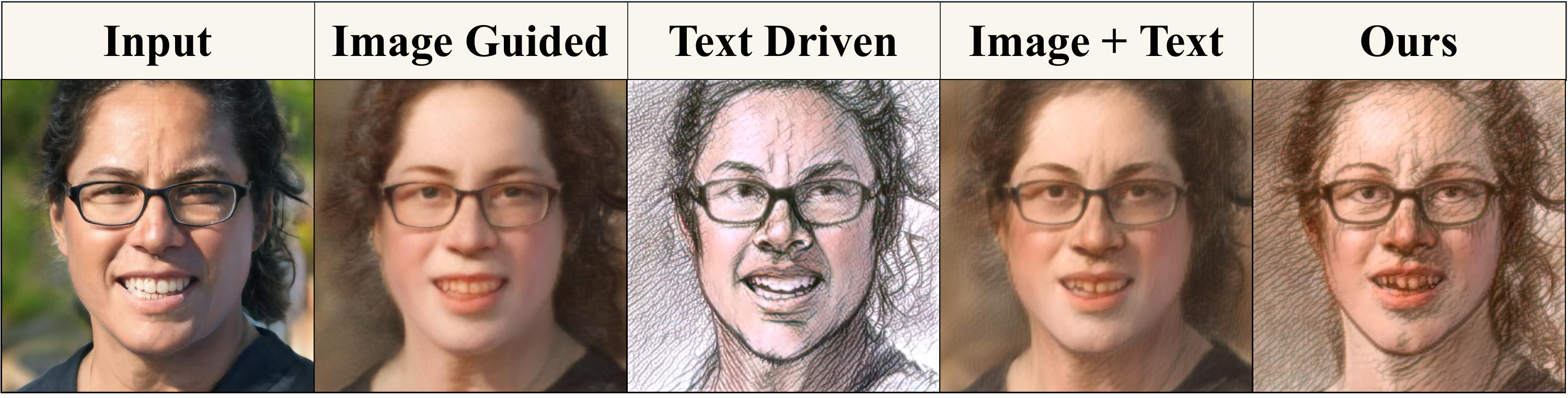}
\end{center}
   \caption{Ablation studies of Image-Text fusion effect in Eq.~\eqref{clip_text}. We compare our strategy with (i) only transfer to the image direction or text direction; (ii) transfer to the middle direction of the image and text with equal $\gamma$ weight. Our proposed method has the best fusion performance.}
\label{fig:ablationbig2}
\end{figure}


\noindent where $\boldsymbol{T}_{target}$ and $\boldsymbol{T}_{source}$ denote the semantic text and input content of the style object, respectively. $\mathbf{I}_{style}$, $\mathbf{I}_{train}$, and $\mathbf{I}_{source}$ are defined in Eq.~\eqref{MIX_clip}. $\mathcal{E}_I$ and $\mathcal{E}_T$ are the image and text encoders of CLIP. $\boldsymbol{L}_{I}$ is used to maintain the image-guided style of the previous stage. $\boldsymbol{L}_{T}$ is used to obtain the text-driven style. $\boldsymbol{L}_{IT}$ is the final CLIP direction loss. As exhibited in Fig. \ref{fig_clip2}, by this means, ITE can achieve Image-Text coupled manipulation to 3D portraits. We use $\gamma$ as a parameter to control the weights guided by text direction $\boldsymbol{L}_{T},$ and image direction $\boldsymbol{L}_{I}$. In the experiments, we found that optimizing the direction of the image and text simultaneously may easily cause an insignificant fusion effect (Fig. \ref{fig:ablationbig2}). Therefore, we design a threshold function to control $\gamma$ so that we only transfer to a certain direction of image or text in each epoch. The expression of  $\gamma$ is defined as follows:

\begin{equation}
\begin{aligned}
\gamma= \left \{
\begin{array}{lr}
    0,                    & if  \quad \Delta{I} \leq \tau   \\
    0,                    & if \quad Rand(1,100) \leq \xi  \\
    1,                                 & otherwise
\end{array}
\right.
 \label{clip_para2}
\end{aligned}
\end{equation}

\noindent where $\Delta{I}$ refers to Eq.~\eqref{clip_text}. we use the $\Delta{I}$ to measure the degree of image-guided stylization, which means the distance of a stylization image from the real photographs in the CLIP space. Because at the beginning of the alternating training approach, $\mathcal{G}_{3d}^{s}(\mathbf{W},\mathbf{P}_{m};\theta)$ in Sec. \ref{section_image} is less stylized, resulting unstable direction of $\boldsymbol{L}_{I}$ and large direction deviation. Therefore, only when distance $\Delta{I}$ exceeds the threshold $\tau$, which means the image is stylized enough, will we transfer to the direction of $\boldsymbol{L}_{I}$. And if the distance $\Delta{I}$ is less than the threshold $\tau \in (0.5,0.75)$, which means the degree of stylization is insufficient, we only perform domain adaptation in the text direction $\boldsymbol{L}_{T}$. In addition, the experiments demonstrate that when training more epochs, $\Delta{I}$ tends to exceed the threshold $\tau$ often. This can lead to over-optimizing the style image-guided direction regardless of the text-driven direction. To handle this issue, we propose another threshold-based text-compensation strategy. We use a random function to generate an integer from 1 to 100. When the integer value is less than $\xi\in (1,100)$, we compensate for transferring in the direction of $\boldsymbol{L}_{T}$. With the threshold techniques $\tau$ and $\xi$, we can automatically perform self-adaptive Image-Text fusion according to the degree of threshold in the CLIP space, which helps to maintain the domain style of image and text.

\begin{figure*}[tp]
\begin{center}
\includegraphics[width=0.95\linewidth]{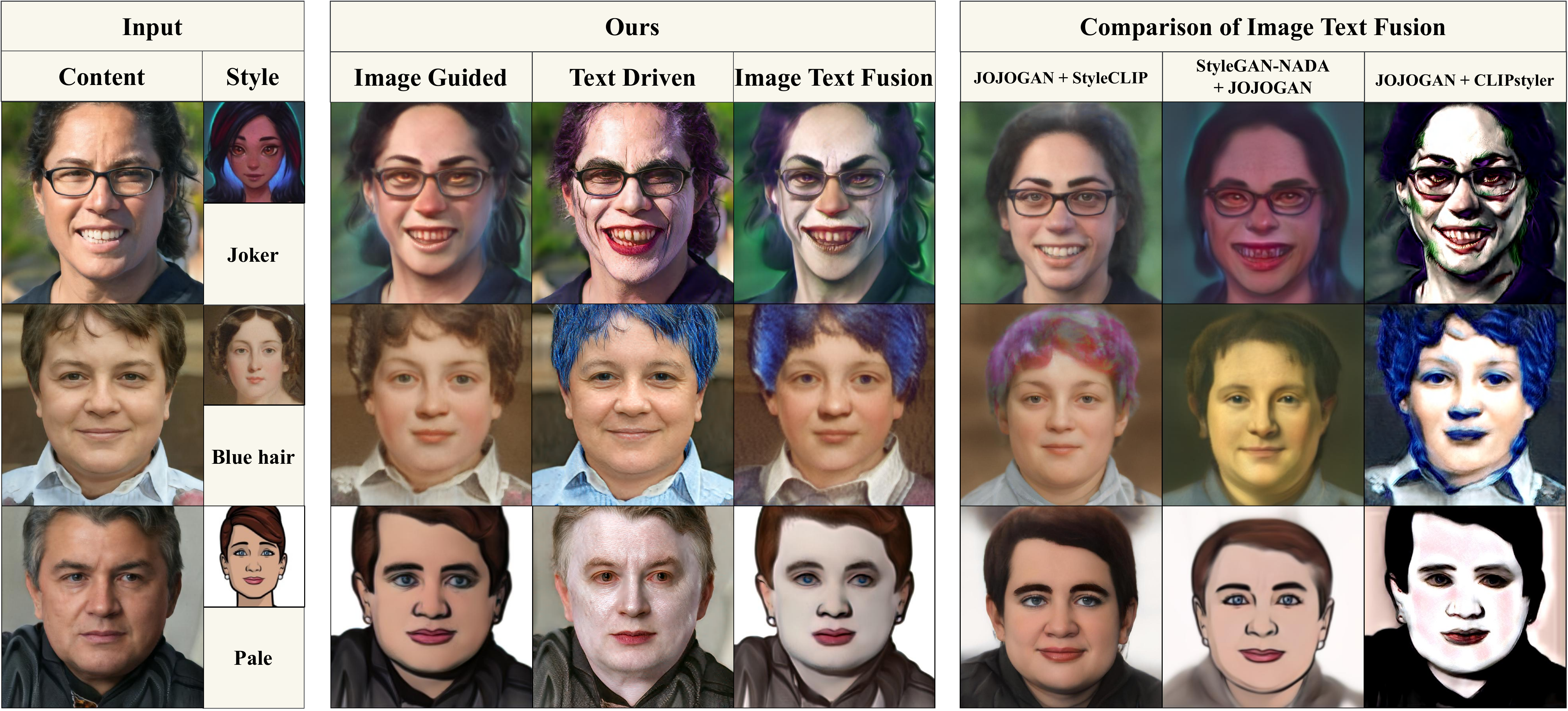}
\end{center}
   \caption{Comparison of domain adaptation effects of Image-Text coupled guidance. We can transfer the precise style from the style image and have flexible text-driven editing ability. While other two-stage methods will lose style of the previous stage.}
\label{fig:zhanshi}
\end{figure*}

\subsection{Alternating Training}\label{section_training}

In this section, we introduce the alternating training mechanism. First, we use the artistic GAN inversion in Eq.~\eqref{equation_w3d} to get $\mathbf{W}_{3d}$ and $\mathbf{P}$. Then start our alternating training mechanism, as shown in Fig. \ref{fig_pipleline_4}. In the first stage, our APT uses Eq.~\eqref{equation_jojo} to train $\mathcal{G}_{3d}^{s}$. In the second stage, our ITE exploits the Eq.~\eqref{clip_text} to train $\mathcal{G}_{3d}^{t}$. $\mathcal{G}_{3d}^{o}$ is always keep frozen. The above two stages are trained alternately.

This alternating training strategy has two advantages: \textbf{(i)} It can avoid the insignificant fusion effect of Image-Text joint training~\cite{StyleCLIPDraw0,styleclipdrawFrans}. Because the threshold function in Eq.~\eqref{clip_para2} can be used to control the $\mathcal{G}_{3d}^{t}$ to only transfer in one direction (image or text) at each epoch. Therefore, a better Image-Text fusion effect can be achieved (Fig.~\ref{fig:ablationbig}).
\textbf{(ii)} It effectively prevents the one-shot image-guided overfitting~\cite{stylegannada} in the CLIP space because $\mathcal{G}_{3d}^{s}$ and $\mathcal{G}_{3d}^{o}$ can generate many different stylized images at each epoch, which help to prevent the $\mathcal{G}_{3d}^{t}$ from overfitting.

\section{Experiments}

\subsection{Implement Details.}  
We use the Adam~\cite{adam} optimizer. The learning is set to $2 \times 10^{-3}$. To speed up the training time, we retain the layer-optimize technique from StyleGAN-NADA~\cite{stylegannada}. Considering the network structure of EG3D, we only optimize the first, third, and fourth modules (SynthesisNetwork, Superresolution, Decoder) of EG3D. Thus, finetuning ITportait on an RTX 3090 for 400 epochs takes about 8 minutes with a batch size of 2 for every case. Furthermore, in Eq.~\eqref{clip_para2}, the approximate value of the threshold $\tau$ is 0.7, and the approximate value of the threshold $\xi$ is 50. $\mathbf{P}_{m}$ in Eq.~\eqref{MIX_clip} is three view angles randomly selected from yaw and pitch. The yaw angles are from $-50^{\circ}$ to $50^{\circ}$ and the pitch angles are from $-30^{\circ}$ to $30^{\circ}$. 

\subsection{Comparsion with Baselines}

\noindent\textbf{Image-Text Coupled Domain adaption.} We conducted comparative experiments to evaluate the effectiveness of our Image-Text coupled (ITE) results. The goal of style fusion is to maintain the reference image's style and the text's editing effect. For a fair comparison, we consider two Image-Text fusion strategies: \textbf{(i)} use the image-guided style transfer method first, and then perform text Editings, such as JOJOGAN~\cite{JoJoGAN0} + StyleCLIP~\cite{Styleclip2} and JOJOGAN~\cite{JoJoGAN0} + CLIPstyler~\cite{CLIPstyler0}. \textbf{(ii)} use text editing first, and then perform image-guided style transfer, such as StyleGAN-NADA~\cite{stylegannada} +JOJOGAN~\cite{JoJoGAN0}. As shown in Fig. \ref{fig:zhanshi}, the result shows that our ITportrait maintains the reference image's style and the text's editing effect best. Specifically, the effect of JOJOGAN + StyleCLIP is not obvious because the style of JOJOGAN cannot be preserved in the text editing stage of StyleCLIP. The effect of JOJOGAN + CLIPstyler has artifacts because CLIPstyler incorrectly optimizes the style of JOJOGAN. The effect of StyleGAN-NADA + JOJOGAN is also not obvious because the stylization of JOJOGAN completely conceals the text editing of StyleGAN-NADA, distorting the style of image and text. In contrast, our method effectively preserves the style and CLIP editing effects thanks to the fusion mechanism ITE in the CLIP space.

\begin{figure}[tp]
\begin{center}
\includegraphics[width=1.0\linewidth]{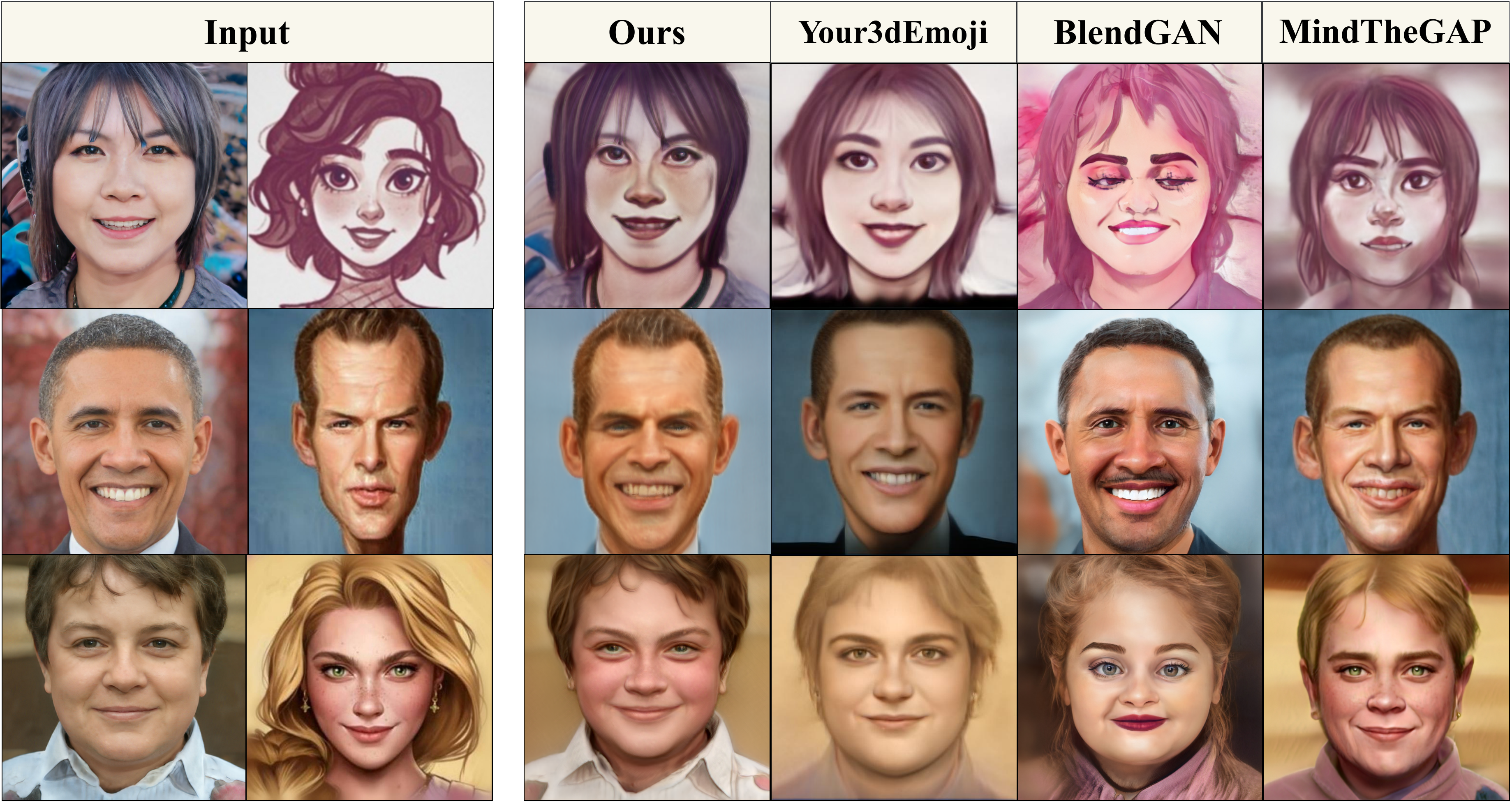}
\end{center}
 \caption{Comparison of image-guided style transfer.} 
\label{fig:img_comparision}
\end{figure}

\noindent\textbf{Image-guided Stylization.} We evaluate our proposed image-guided one-shot stylization method APT. For image-guided style transfer tasks, a high-quality transfer effect refers to the style that can be transferred while maintaining the Identity of the original portrait. We selected these SOTA portrait style transfer methods for comparison, including Your3dEmoji~\cite{Your3dEmoji0}, BlendGAN~\cite{jojogan21}, MindtheGAP~\cite{jojogan41}. The results are illustrated in Fig. \ref{fig:img_comparision}, and our effect is the best. Specifically, BlendGAN can only capture the general color of the style, and the specific portrait details are insufficient. Your3dEmoji can achieve higher-quality stylization details. But their wrinkle geometry deformation is inferior, such as for the caricature example (Line 2). The effect of MindtheGAP is the closest to ours. But MindtheGAP needs about 15 minutes to train, while our ART only needs 4 minutes of finetuning.

\begin{figure}[tp]
\begin{center}
\includegraphics[width=1.0\linewidth]{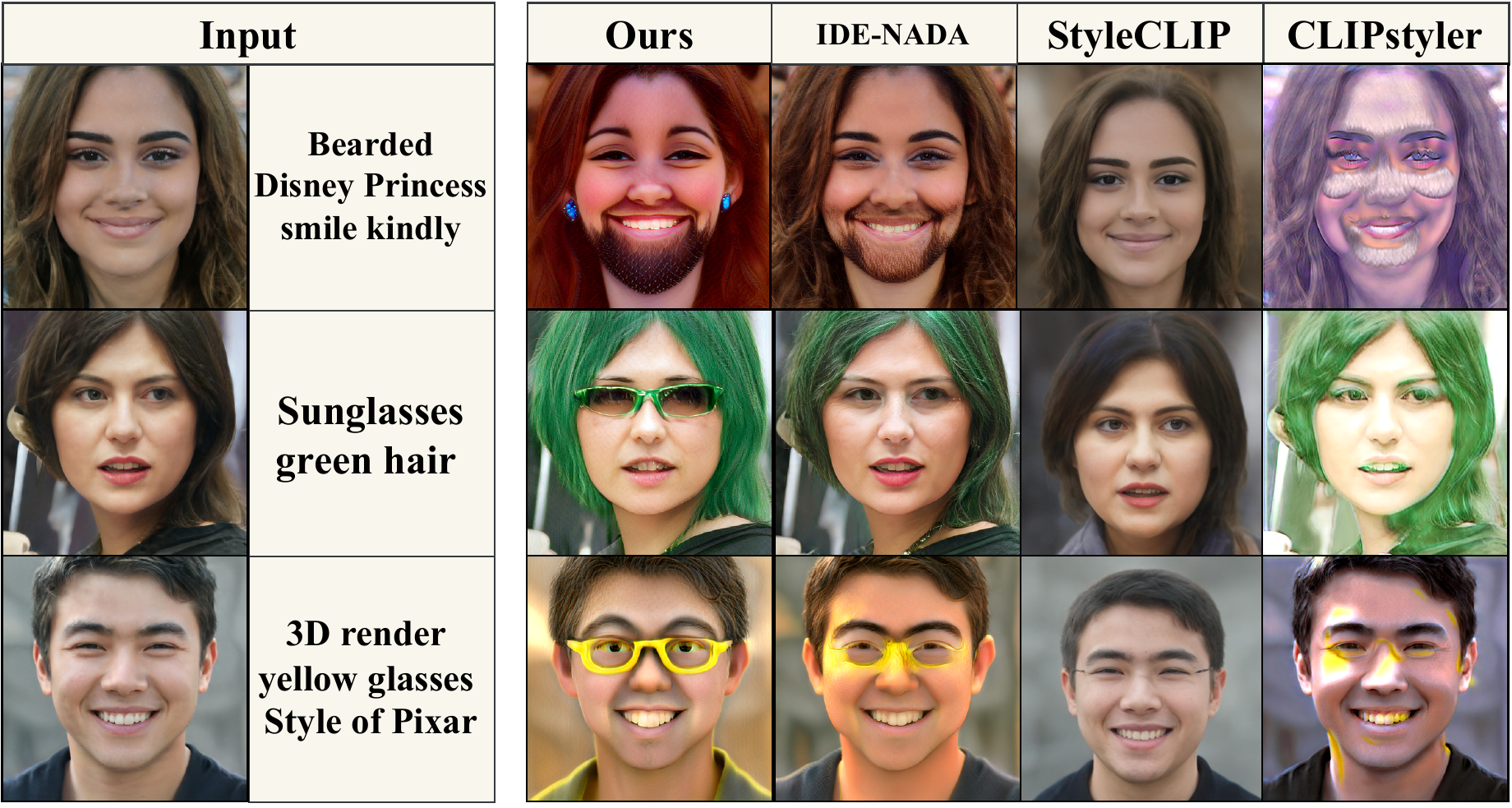}
\end{center}
   \caption{Comparison of text-driven manipulation.} 
\label{comparison-text-driven2}
\end{figure}

\noindent\textbf{Text-driven Manipulation.} ITportrait supports the manipulation of portraits with only text guidance. We compare our method with the SOTA text-driven manipulation methods including, CLIPstyler~\cite{CLIPstyler0}, StyleCLIP~\cite{Styleclip2}, and IDE-NADA (IDE-3D~\cite{40Ide3d} + StyleGAN-NADA~\cite{stylegannada}). Because StyleCLIP is good at face editing, while IDE-NADA and CLIPStyler are good at domain transfer. Therefore, for a fair comparison, we use long text descriptions containing face edits and cross-domain information. As shown in Fig.~\ref{comparison-text-driven2}, our method achieves more visually pleasing domain adaptation results. This stems from the fact that we construct 3D multi-view supervision. We effectively supervise the portrait's side and achieve the CLIP enhancement effect. Although IDE-NADA is aimed at 3D portraits, it only employs 2D CLIP enhancement methods~\cite{styleclipdrawFrans,CLIPstyler0}. This leads to poor domain adaptation results for some text descriptions. For the example of 'wearing glasses', the effect of IDE-NADA is not obvious (Lines 2 and 3).

\noindent\textbf{Quantitative Analysis.} 
Quantitative analysis of style transfer and cross-domain editing is challenging due to the absence of GroundTruth. Conducting a user study is the most common way to evaluate different style transfer methods. We use a similar approach to~\cite{stylegannada,CLIPstyler0}. Specifically, we establish four sets of images. Users have unlimited time to rate their preferences from 1 to 5 (five being the best and one being the worst). All images from each group are presented side-by-side in random order. Each group contains comparisons between our method and other methods, including JOJOGAN~\cite{JoJoGAN0} + CLIPstyler~\cite{CLIPstyler0},
StyleGAN-NADA~\cite{stylegannada} + JOJOGAN~\cite{JoJoGAN0}, and
JOJOGAN~\cite{JoJoGAN0} + StyleCLIP~\cite{Styleclip2}. As shown in Tab. \ref{table:user2}, our method is the most popular. Because our ITportrait most effectively retains the image-guided and text-driven styles, thus achieving the best image-text fusion adaptation effect. Please refers to the supplement for more details about user studies.
\vspace{3.7mm}

\begin{table}[tbp]
\begin{center}
\centering
  \footnotesize
  \caption{Quantitative analysis of user study.}
		\begin{tabular}{c|ccc}
        \toprule
      \textbf{Methods} & \textbf{Image-Style $\uparrow$} & \textbf{Text-Style $\uparrow$} & \textbf{Content $\uparrow$} \\ \midrule
      JOJOGAN + StyleCLIP & 1.119 & 1.343  & \textbf{3.914} \\
       StyleGAN-NADA + JOJOGAN & 3.716 & 1.740 & 3.072 \\
       JOJOGAN + CLIPstyler & 3.466 & 3.986  & 2.867 \\
      Ours & \textbf{4.775} & \textbf{4.842}  & 3.742 \\
     \bottomrule
    \end{tabular}
    \label{table:user2}
\end{center}
\end{table}

  \begin{figure}[tp]
\begin{center}
\includegraphics[width=0.9\linewidth]{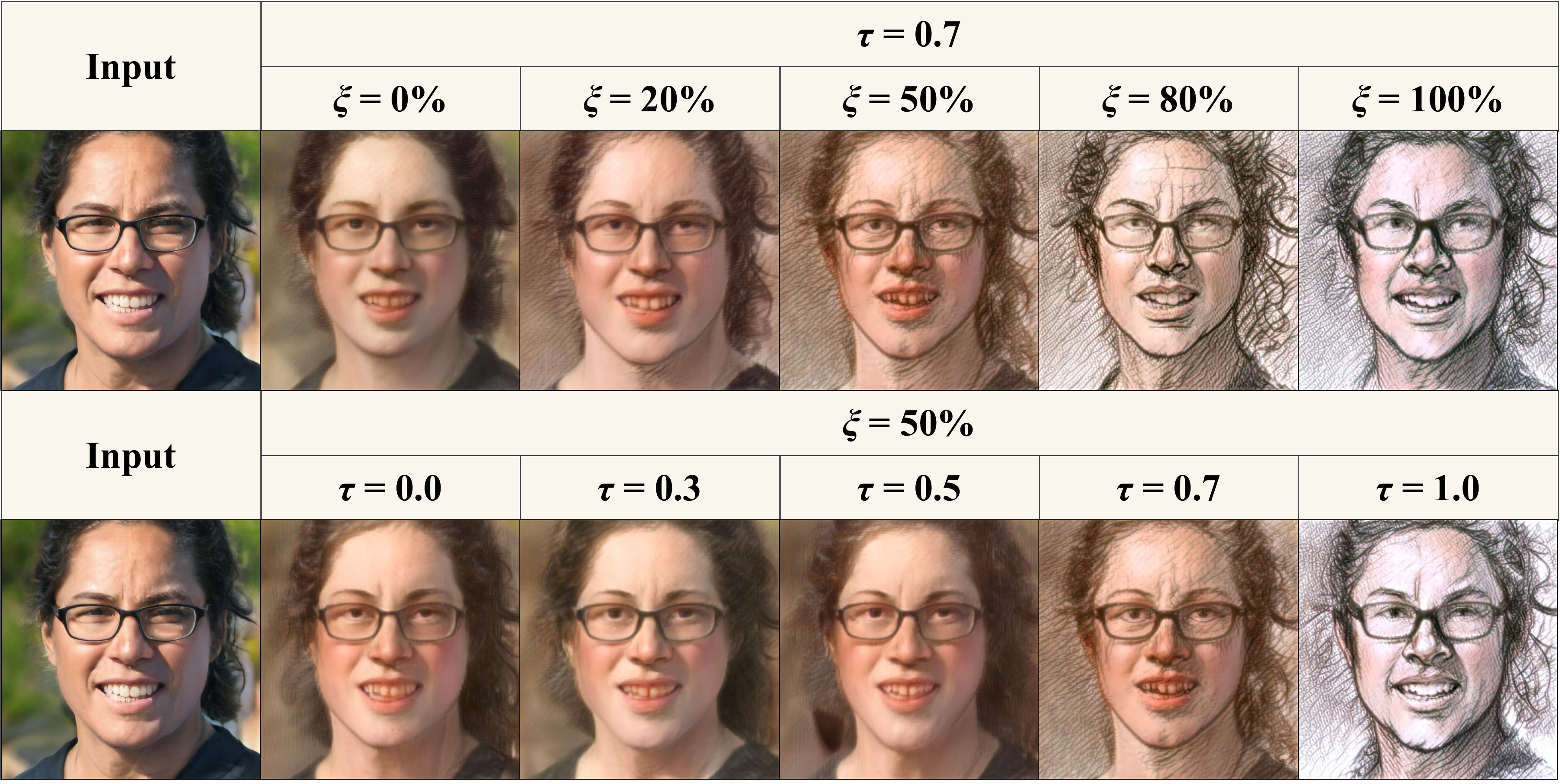}
\end{center}
   \caption{Ablation studies of threshold function in Eq.~\eqref{clip_para2}. }
\label{fig:ablationbig}
\end{figure}


\subsection{Ablation Studies}

\noindent\textbf{Pose estimation.} We conduct ablation experiments to evaluate the effectiveness of our APT that uses photo-realistic portraits for pose estimation (Eq.~\eqref{equation_w2d}). Specifically, we compare: \textbf{(i)} not use APT initialization, directly optimizing with Eq.~\eqref{equation_w3d}  and \textbf{(ii)} use APT initialization and then optimize with Eq.~\eqref{equation_w3d}. We select 30 art images to compare the estimation of GAN inversion accuracy. The results are listed shown in Tab. \ref{tabpose}. It can be observed that using APT can effectively improve our quality. Because the 3D GAN inversion~\cite{ko20233d,yin20223d} of artistic portraits, pose $\mathbf{P}$ and $\mathbf{W}_{3d}$ are randomly initialized and simultaneously optimized may lead to artifact results. We construct real photographs as the initialization of the pose, which can ease the difficulty of pose optimization, thereby improving the quality of the one-shot stylization. More details are shown in the supplement.

\noindent\textbf{Image-Text Fusion Strategy.}
We conduct ablation experiments to prove the effectiveness of our self-adaptive threshold function (Eq.~\eqref{clip_para2}) for the ITE in CLIP space. The threshold $\tau$ is used to transfer to the image direction when the $\mathcal{G}_{3d}^{s}$ is stylized enough. The threshold $\xi$ can be used to maintain the Text style and prevent the excessive transfer to the image direction. Specifically, we compare our ITE with \textbf{(i)} do not use the threshold function Eq.~\eqref{clip_para2} at all, \textbf{(ii)} only use the threshold $\tau$, and \textbf{(iii)} only use the threshold $\xi$. We keep other parameters constant and only change the different thresholds mentioned above. In addition, we also explored the impact of different thresholds $\xi$ and $\tau$ on the fusion results. As shown in Fig. \ref{fig:ablationbig}, our threshold method achieves the best fusion effects. It can be seen that if the threshold function is not used, simultaneously optimizing CLIP and image loss has no obvious effect because there is a contradiction between image guidance and text guidance. Threshold $\tau$ and $\xi$ can be effectively used to preserve the orientation of the image and the text, respectively.

\begin{table}[tbp]
  \centering
  \footnotesize
  \caption{Ablation analysis of pose initialization. We use Eq.~\eqref{equation_w3d} to get the GAN inversion result of the art image to evaluate the pose estimation accuracy. The comparison includes (i) using random weight \cite{ko20233d,yin20223d} without initialization. and (ii) using Eq.~\eqref{equation_w2d} as initialization.} 
  \begin{tabular}{cccccc}
      \toprule
      Method    & MSE $\downarrow$   & SSIM  $\uparrow$   & PSNR $\uparrow$ & LPIPS  $\downarrow$ & ID  $\uparrow$  \\
      \midrule
      w/o pose init  & $0.0947$ & $0.4612$ & $10.2348$ & $0.2701$ & $0.0294$  \\
     w/ pose init & \textbf{$0.0052$} & \textbf{$0.7884$} & \textbf{$22.7444$} & \textbf{$0.1058$} & \textbf{$0.8372$} \\
      \bottomrule
  \end{tabular}
\label{tabpose}
\end{table}

  \begin{figure}[tp]
\begin{center}
\includegraphics[width=0.9\linewidth]{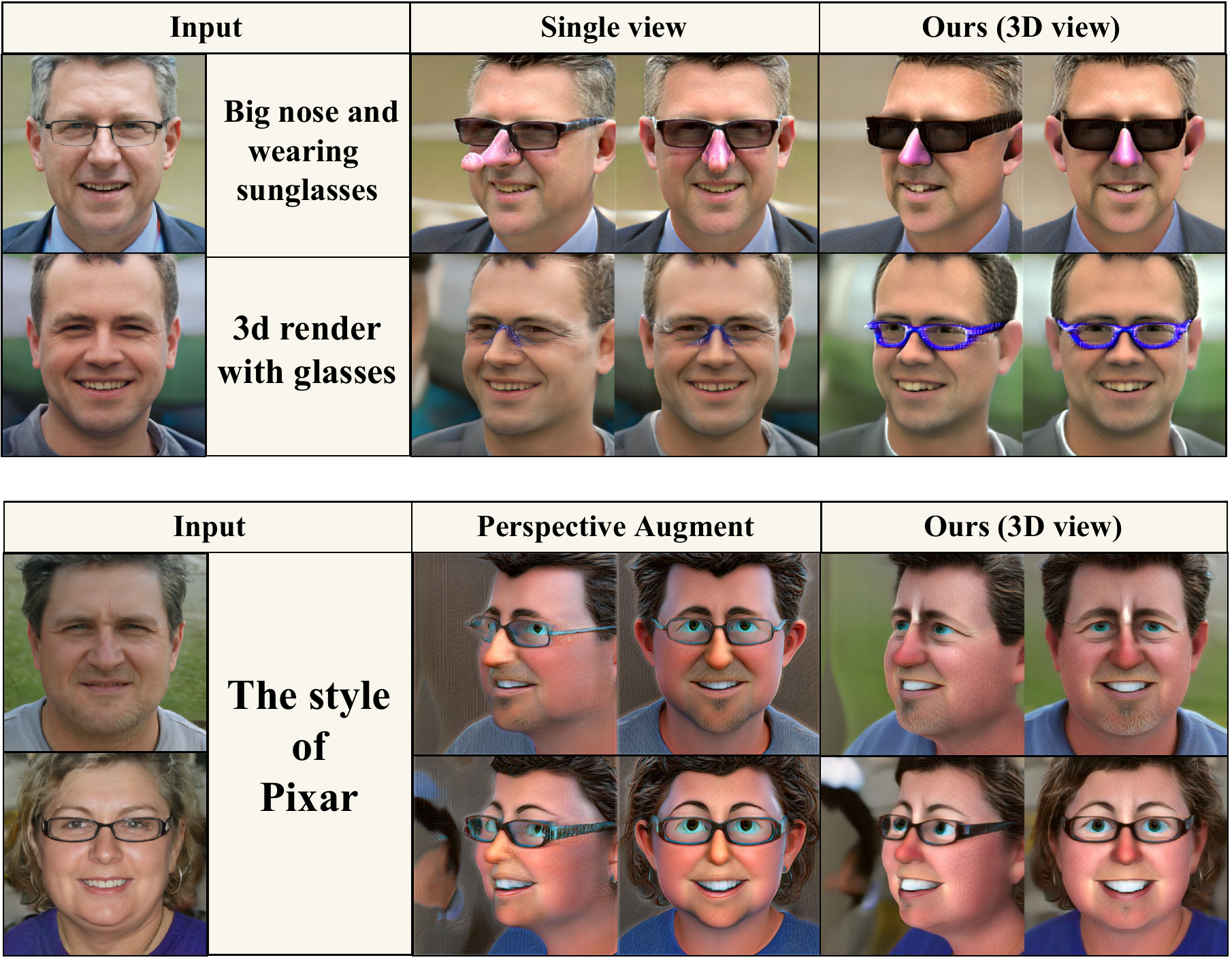}
\end{center}
   \caption{Ablation studies for 3D view augment in Eq.~\eqref{MIX_clip}. If only the front single view~\cite{sun2022next3d,40Ide3d} is used to supervise, the geometry of the nose will collapse (Line 1), and the effect of 'wearing glasses' will not be obvious (Line 2). The perspective augment~\cite{StyleCLIPDraw0,CLIPstyler0} will misinterpret the wrinkles on the eyes as the effect of wearing glasses (Line 3) or cause blurry glasses on the side (Line 4).}
\label{fig:ablation2}
\end{figure}

\noindent\textbf{3D View Augment.} 
 We conduct ablation experiments to demonstrate the effectiveness of our 3D view augments. Specifically, we compare with the original single-view~\cite{sun2022next3d,40Ide3d} and perspective augment~\cite{StyleCLIPDraw0,CLIPstyler0}. For fairness, we keep other parameters the same and only change the number of supervised views. Specific Results As shown in Fig. \ref{fig:ablation2}, our method works best. There is a portrait geometric collapse in single-view supervision (Lines 1,2). The CLIP perspective augments method can misinterpret the wrinkles on the eyes as the effect of wearing glasses (Lines 3) or cause blurry glasses on the side (Line 4).
In summary, although these augment methods positively affect the front portrait, the side of the portrait is blurred. In contrast, our method is free from distortions in all views. Because we can effectively supervise the side faces.

\begin{figure}[tp]
\begin{center}
\includegraphics[width=0.95\linewidth]{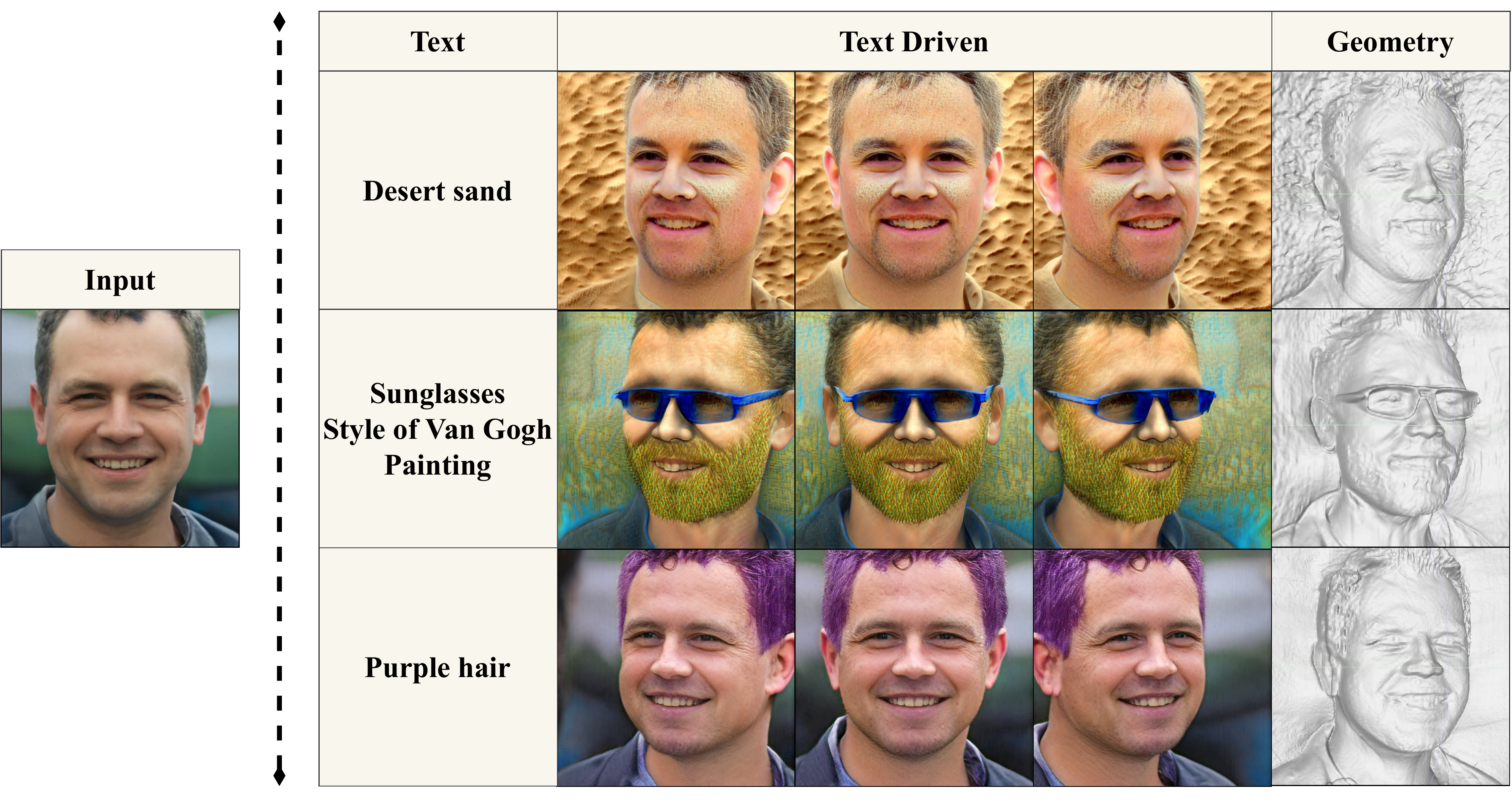}
\end{center}
   \caption{Application of new perspective synthesis.}
\label{fig:newview}
\end{figure}

\begin{figure}[tp]
\begin{center}
\includegraphics[width=0.95\linewidth]{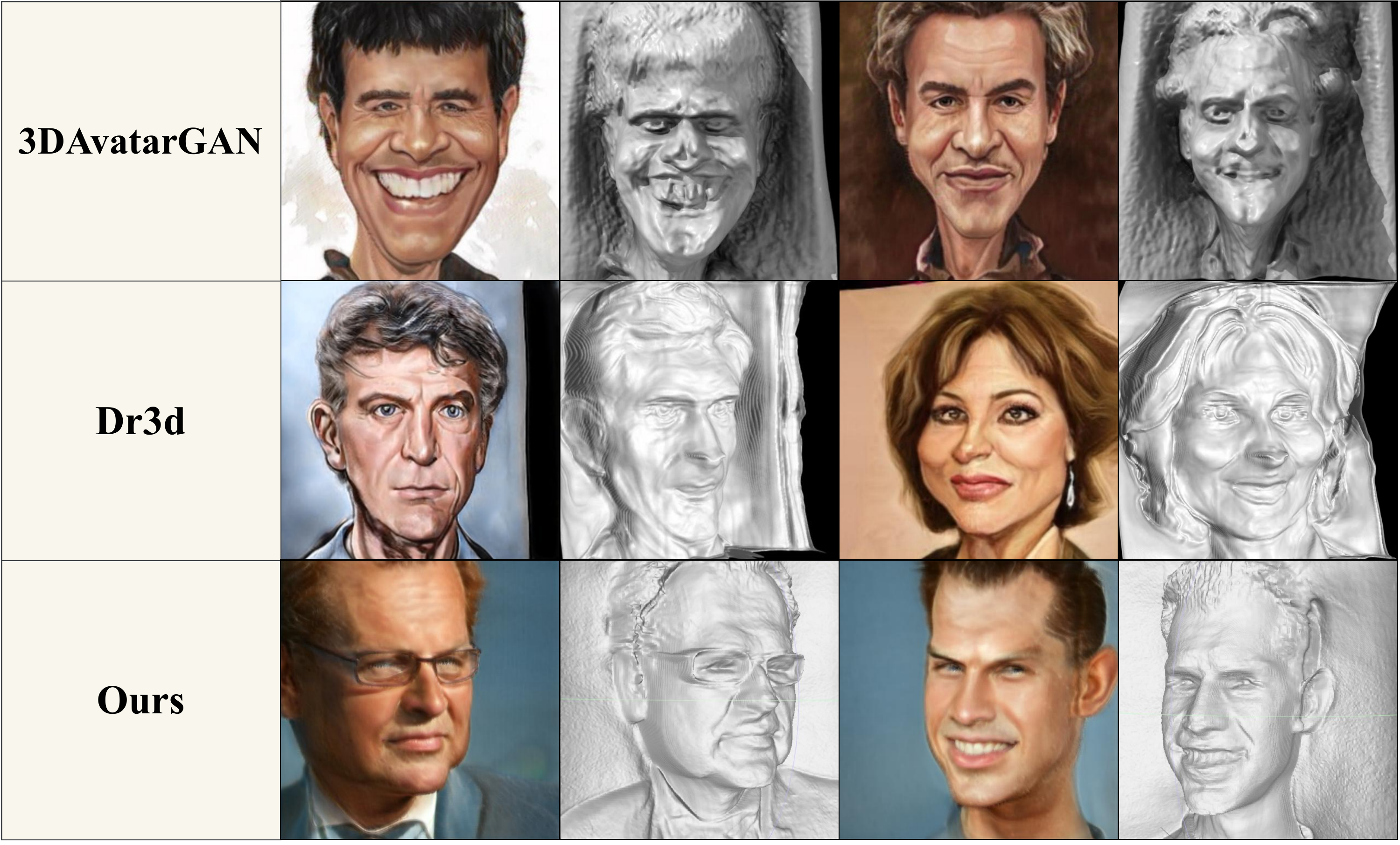}
\end{center}
   \caption{Applicaiton of 3D portrait domain adaption. We compare our method with 3DAvatarGAN~\cite{3DAvatarGAN} and Dr.3d~\cite{dr3d} in the caricature domain.}
\label{fig:3ddomain}
\end{figure}

\subsection{Application}

\noindent\textbf{Novel View Synthesis.} As depicted in Fig. \ref{fig:newview}, our ITportrait can support novel view synthesis of images from cross-domain drawings. The main reason is that ITportrait aggregates image-guided style and text-driven manipulation in the CLIP space while adding 3D multi-view supervision. As shown in the ablation study in Fig. \ref{fig:newview}, the proposed ITportrait effectively prevents the appearance of portrait side artifacts and geometric collapse. Thus, we can achieve better high-quality perspective synthesis than the previous  methods~\cite{sun2022next3d,40Ide3d,StyleCLIPDraw0,CLIPstyler0}.

\noindent\textbf{3D Portrait Domain Adaption.} Our method supports 3D portrait domain adaptation with limited data (\emph{i.e.,} a text or a single image). Fig. \ref{fig:3ddomain} shows the 3D portrait domain adaptation effect of Dr.3d~\cite{dr3d} and 3DAvatarGAN~\cite{3DAvatarGAN} in the caricature domain. However, they need a dataset of the corresponding art domain~\cite{Resolution2, Pastiche} and train from scratch, which is very time-consuming (training for about 4 hours). Furthermore, these large-scale corresponding domain datasets are tedious and labor-intensive to collect. In contrast, With 8 minutes of fine-tuning and a single image or text as guidance, ITportrait supports 3D portrait domain adaptation.

\begin{figure}[tp]
\begin{center}
\includegraphics[width=0.95\linewidth]{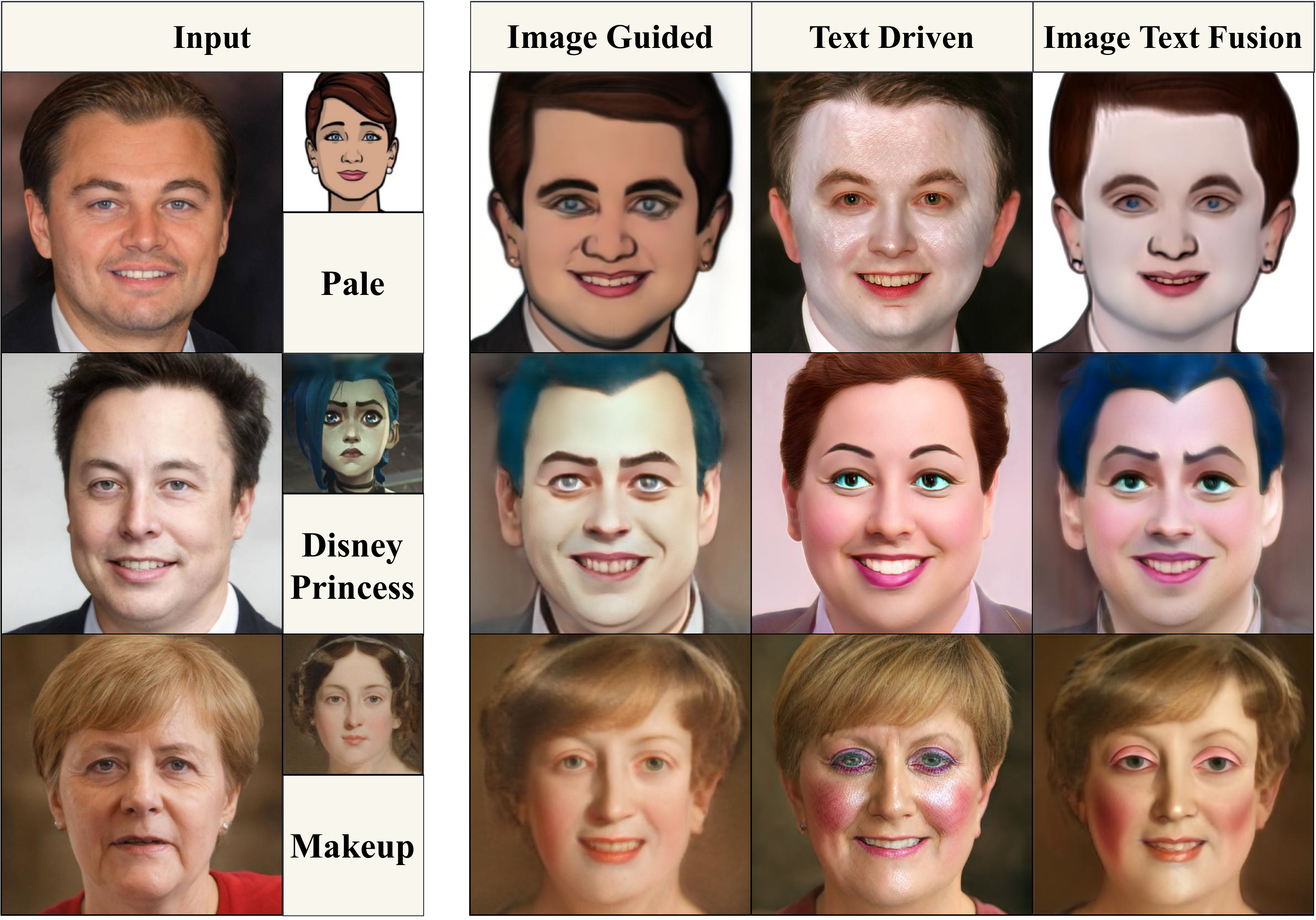}
\end{center}
   \caption{Application of photo-realistic portrait editing.}
\label{fig:inversion}
\end{figure}

\begin{figure}[tp]
\begin{center}
\includegraphics[width=0.95\linewidth]{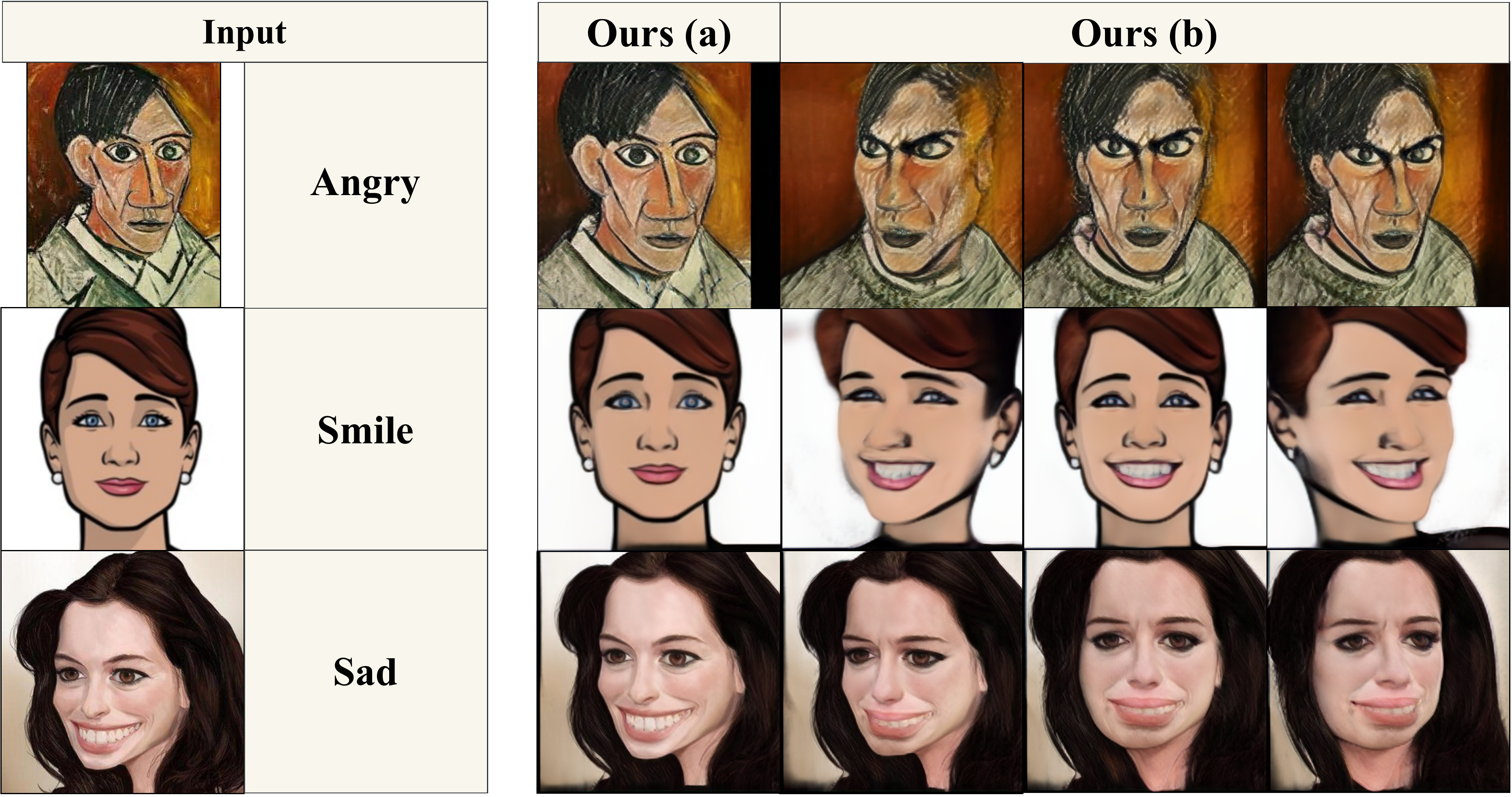}
\end{center}
   \caption{Application of artistic drawing portrait editing. Ours (a) is the result of 3D GAN inversion. Ours (b) is the result of text-driven view-consistent editing.}
\label{fig:art3d}
\end{figure}



\noindent\textbf{View-Consistent Portrait Editing.} Our ITportrait can achieve view-consistent portrait editing. As illustrated in Fig. \ref{fig:inversion}, our ITportrait can implement image-guided, text-driven, and Image-Text coupled domain adaptation for photo-realistic portraits. As shown in Fig.~\ref{fig:art3d}, we also support art-drawing portrait editing. Specifically, we can use APT for two-stage 3D GAN inversion, similar to PTI~\cite{roich2022pivotal}. We first get the pivot latent code $\mathbf{W}_{3d}$ and $\mathbf{P}$ (in Eq. ~\eqref{equation_w3d}). Then we fine-tune $\mathcal{G}_{3d}$ using Eq.~\eqref{equation_jojo}. This way, APT can realize high-quality 3D GAN inversion for both photo-realistic and art-drawing portraits. Then we implement editing using our ITE approach. Please refers to our supplement for more details.

\section{Conclusions}

In this paper, We propose ITportrait, a novel framework for 3D portrait Image-Text coupled domain adaptation. To the best of our knowledge, this is the first work to study Image-Text coupled manipulation for 3D portrait domain adaptation, which can offer better operability of artistic portraits. We design a two-stage alternating training strategy for image and text fusion in the CLIP space. Specifically, In the first stage, we design a one-shot 3D portrait style transfer method (APT). Our APT constructs a paired photo-realistic portrait to accurately estimate the pose of artistic images for high-quality one-shot stylization. In the second stage, we present an Image-Text embedding (ITE) strategy. ITE includes 3D supervision and a threshold function to self-adaptively achieve controllable fusion direction control. Comprehensive experiments demonstrate that ITportrait achieves state-of-the-art (SOTA) results and supports wide downstream applications.

\newpage
\bibliographystyle{ACM-Reference-Format}
\bibliography{sample-sigconf}


\end{document}